\documentclass[prd,twocolumn,nofootinbib,notitlepage,aps,tightenlines,preprintnumbers,amsmath,amssymb,amsfonts,showpacs,superscriptaddress]{revtex4-1}

\RequirePackage[colorlinks=true
,urlcolor=blue
,anchorcolor=blue
,citecolor=blue
,filecolor=blue
,linkcolor=blue
,menucolor=blue
,linktocpage=true
,pdfproducer=medialab
,pdfa=true
]{hyperref}

\usepackage{amsmath,amssymb,amsthm,amsfonts}
\usepackage{graphicx,tabularx}
\usepackage{float}
\usepackage{multirow}  
\usepackage{hyperref}
\usepackage{cancel}
\usepackage{comment}
\usepackage{physics}

\usepackage{cleveref}
\crefname{section}{Sec.}{Secs.}
\crefname{figure}{Fig.}{Figs.}
\crefname{equation}{Eq.}{Eqs.}
\crefname{appendix}{Appendix}{Appendices}

\newcommand{\be}{\begin{equation} \begin{aligned}}
\newcommand{\ee}{\end{aligned} \end{equation}}

\usepackage{xcolor}
\usepackage{lineno}

\RequirePackage[normalem]{ulem} 
\newcommand{\changed}[2]{{\protect\color{red}\sout{#1}}{\protect\color{blue}\uwave{#2}}}

\newcommand{\orcid}[1]{\begingroup
  \hypersetup{hidelinks}\href{https://orcid.org/#1}{\includegraphics[width=10pt]{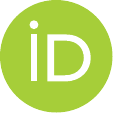}} \endgroup}

\begin{document}

\title{
The 
LHC as a TeV Muon Beam Dump: 
\\
Muonphilic Scalars at FASER
}

\author{Roshan Mammen Abraham \orcid{0000-0003-4678-3808}}
\email{rmammena@uci.edu}
\affiliation{Department of Physics and Astronomy, University of California, Irvine, CA 92697 USA\\[0.1cm]}

\author{Max Fieg \orcid{0000-0002-7027-6921}}
\email{mfieg@uci.edu}
\affiliation{Department of Physics and Astronomy, University of California, Irvine, CA 92697 USA\\[0.1cm]}

\begin{abstract}
The FASER experiment was designed to study long-lived dark sector particles and neutrinos traveling in the forward direction at the LHC. 
Neutrinos are predominantly produced from meson decays, which also result in an intense energetic flux of muons in the forward direction regularly observed by FASER. 
So far, these muons are treated only as backgrounds to neutrino and new physics studies, and extensive effort is required to suppress them. 
In this study, we consider the opposite scenario and use muons produced in the forward direction to produce new muonphilic scalars, which can then be searched for at the FASER detector. 
To minimize the backgrounds for this search, we make use of an upgraded preshower component, which is expected to be installed at FASER before the end of Run 3, and is capable of spatially resolving two energetic photons. 
We find that FASER, and its upgrade, FASER2
can probe currently unconstrained regions of parameter space, including regions that can potentially explain the $(g-2)_{\mu}$ anomaly.
This highlights the physics opportunities that the intense TeV muon beam at the LHC can bring.

\end{abstract}

\maketitle


\section{Introduction}

The forward direction at the Large Hadron Collider (LHC) was recognized as early as 1984 to be an important source of neutrino and muon beams~\cite{DeRujula:1984pg,DeRujula:1992sn}. This idea was resurrected and eventually realized as the FASER experiment~\cite{Feng:2017uoz} to study high energy neutrinos and search for dark sector particles. FASER~\cite{FASER:2018ceo,FASER:2018bac} and the associated neutrino detector FASER$\nu$~\cite{FASER:2019dxq,FASER:2020gpr} were installed in 2021 and have been taking data since then. An impressive collection of neutrino~\cite{FASER:2021mtu,FASER:2023zcr,FASER:2024hoe,FASER:2024ref}, and beyond the Standard Model (BSM) ~\cite{FASER:2023tle,FASER:2024bbl} physics results have already been reported.\footnote{ SND@LHC~\cite{SNDLHC:2022ihg}, has also reported the detection of collider neutrinos~\cite{SNDLHC:2023pun}.} The TeV-scale muon beam in the forward direction at the LHC has been less studied in the literature.

Proton collisions at the LHC produce a large flux of hadrons in the forward direction, and their subsequent decay results in a beam of muons. Muons can also be produced from particle interactions with downstream LHC infrastructure, but this is a subleading contribution.
These muons will pass through several parts of the surrounding LHC infrastructure, and a large number of them 
will pass through FASER, which is located on the beam-axis far from the ATLAS interaction point (IP). 
An estimated $2\times10^{9}$ muons will pass through the detector during Run 3 of the LHC~\cite{FASER:2021mtu,FASER:2018bac,FASER:2019dxq,FASER:2020gpr}. Even though a majority of the flux is at ${\cal O}(100)$ GeV, there still remains a significant flux at energies larger than a TeV (see Fig.~2 in Ref.~\cite{FASER:2021mtu}).
These muons are a major source of background for neutrino and dark sector searches at FASER, and substantial effort is put into suppressing them.
However, it is interesting to consider the new physics opportunities that might be achieved with this muon beam.
In this work, we explore the possibility that this intense and energetic muon beam can be used to study new muonphilic scalar particles
at FASER.
Although we focus on a specific model here, this work underscores the versatility of the LHC, particularly when combined with forward detectors, as together they effectively function as a TeV-scale muon beam dump.

The muonphilic scalar model we study is a simple model with a number of compelling motivations. 
This model provides an attractive minimal explanation to the observed discrepancy between the measured value of muon's magnetic moment and the expected value~\cite{Jegerlehner:2009ry}.
The recent measurements of $a_{\mu},~\equiv(g-2)_{\mu}/2,$ at the E821 experiment~\cite{Muong-2:2006rrc} and the Fermilab Muon g-2 Experiment~\cite{Muong-2:2021ojo,Muong-2:2023cdq} observed a discrepancy with the Standard Model (SM) prediction~\cite{Blum:2013xva,Aoyama:2020ynm}. The world experimental average deviates from SM prediction by 4.2$\sigma$.\footnote{The theoretical prediction~\cite{Colangelo:2022jxc} is based on calculations that include the contribution of the hadronic vacuum polarization (HVP) inferred from data. It is worth noting that recent calculations of the HVP contribution via lattice QCD methods~\cite{Borsanyi:2020mff,RBC:2023pvn,FermilabLatticeHPQCD:2023jof} differ from the data-driven method, bringing the theoretical prediction 
in better agreement with the measurement. The data-driven techniques extract the HVP contribution from a number of experiments including BaBar~\cite{BaBar:2009wpw,BaBar:2012bdw}, CMD-3~\cite{CMD-3:2023alj} and KLOE~\cite{KLOE-2:2017fda}.
}
The muonphilic scalar is also interesting as a portal to the dark sector capable of accounting for the observed relic abundance of dark matter~\cite{Batell:2009di,Abdughani:2021oit,Ghorbani:2023kke,Calibbi:2018rzv,Krnjaic:2022ozp,Aliberti:2025beg}. The muonphilic model can also be seen as low-energy manifestations of flavor-specific UV-complete theories~\cite{Batell:2017kty,Batell:2021xsi}. 
Furthermore, a displaced detector like FASER is particularly well-suited to study this model for scalar masses below the dimuon mass threshold because the scalar is long-lived due to its loop-suppressed decay to two photons. 

This paper is organized as follows. We introduce our muonphilic scalar model in \cref{sec:model} and the experimental setup is described in \cref{sec:detectors}. In \cref{sec:signal} we discuss muonphilic scalar production, the signal at FASER, and associated backgrounds. We present our results in \cref{sec:results} and in \cref{sec:conclusion} we conclude. In \cref{sec:appendix}, we consider how some additional experimental systematics may impact the discovery potential.

\section{Model}\label{sec:model}
We consider a minimal model with a scalar $S$ coupling only to muons. 
Scalars that also couple to electrons
have tight constraints coming from electron beam experiments~\cite{Davier:1989wz,Liu:2020qgx,Bjorken:1988as,Marsicano:2018vin,BaBar:2020jma} but these constraints are significantly weakened when considering a scalar that only couples to muons.
The effective Lagrangian is given by
\begin{equation}\label{eq:lagrangian}
    \mathcal{L}\supset \frac{1}{2}\left(\partial_{\mu}S\right)^2 - \frac{1}{2}m_S^2S^2-g_{S}S\Bar{\mu}\mu~,
\end{equation}
where $m_S$ is the mass of the scalar, and $g_S$ is the coupling constant. 
This interaction is not invariant under SM symmetries but can be derived from a dimension-5 operator of the form $SH^{\dagger}Ll_R$, where $L,~l_R$ are the lepton doublet and singlet respectively, $H$ is the SM Higgs doublet, and $S$ is our new scalar. This effective Lagrangian can be UV completed in many ways, see e.g. Refs.~\cite{Batell:2017kty,Krnjaic:2019rsv,Capdevilla:2021kcf,Jiang:2024cqj}.

The model in~\cref{eq:lagrangian} is one of the most minimal models that can solve the $(g-2)_{\mu}$ anomaly. This scalar can contribute to the muon's anomalous magnetic moment~\cite{Jackiw:1972jz,Leveille:1977rc,Lindner:2016bgg}, with a contribution expressed as:
\begin{equation}\label{eq:g-2_contribution}
    a_{\mu}=\frac{g_{S}^2}{8\pi^2}\int^1_0dx\frac{\left(1+x\right)\left(1-x\right)^2}{\left(1-x\right)^2+x\left(m_S/m_{\mu}\right)^2}~. 
\end{equation}
In particular, a light scalar with $m_S\sim100$ MeV and $g_{S}\sim5\times10^{-4}$ has the correct contribution to $a_{\mu}$ to explain the $(g-2)_{\mu}$ anomaly. 

Muonphlilic scalars have been searched for at a number of experiments. A strong constraint for $m_S>2m_{\mu}$ has been placed by BaBar~\cite{BaBar:2016sci,Capdevilla:2021kcf,Krnjaic:2019rsv,Chen:2017awl}, while ongoing measurements at Belle II~\cite{Capdevilla:2021kcf} and SpinQuest~\cite{Apyan:2022tsd,Forbes:2022bvo}, and future measurements at the HL-LHC~\cite{Capdevilla:2021kcf},
will either discover or place bounds on muonphilic scalars above the dimuon mass threshold, $2m_{\mu}<m_S<80$ GeV~\cite{Harris:2022vnx}.
For $m_S<2m_{\mu}$ constraints from SN1987~\cite{Marsicano:2018vin,Croon:2020lrf,Rella:2022len} E137~\cite{Bjorken:1988as,Marsicano:2018vin}, NuCal\cite{Blumlein:1991xh}, MiniBooNE, and MicroBooNE~\cite{Cesarotti:2023udo} dominate the small coupling region of the parameter space, up to $m_S\lesssim150$ MeV.
Interestingly, these searches don't rule out the $(g-2)_{\mu}$ favored parameter space below the dimuon mass,
with solutions for scalar masses greater than 150 MeV not yet excluded. 
See Fig.~2 in Ref.~\cite{Harris:2022vnx} for a summary of the existing bounds.

Given the existing and upcoming constraints from past and ongoing experiments, 
we explore in detail the muonphilic scalar model for $m_S < 2m_{\mu}$.
The effective Lagrangian in \cref{eq:lagrangian} induces a dimension-5 coupling between $S$ and the photon field, $SF^{\mu\nu}F_{\mu\nu}$, via a muon loop~\cite{Blinov:2024gcw}. Below the dimuon mass threshold, $S$ decays to two photons through this vertex.\footnote{Another possible decay channel for $S$ is via an off-shell muon. This channel, $S\rightarrow \mu \mu^*(e\nu_e\nu_\mu$), is suppressed by $G_F^2$ and is small in comparison to the diphoton channel, and so it can be ignored.} 
The decay width to two photons in this model is given by~\cite{Chen:2017awl,Marsicano:2018vin,Krnjaic:2019rsv,Rella:2022len}
\begin{equation}\label{eq:decay_width}
    \Gamma=\frac{\alpha^2 g_{S}^2 m_S^3}{64\pi^3 m_{\mu}^2}\abs{\tau\left[1+(1-\tau)f(\tau)\right]}^2~,
\end{equation}
where $\tau=4m_{\mu}^2/m_{S}^2$ and $\alpha$ is the fine structure constant. For $\tau>1~(m_S<2m_{\mu})$, we have $f(\tau)=\arcsin^2(\tau^{-1/2})$. As this interaction is loop-suppressed, it naturally renders $S$ as a long-lived particle which can be searched for at far detectors. For a muonphilic scalar with energy $E_S$, this gives a decay length of~\cite{Chen:2017awl}
\begin{equation}\label{eq:decay_length}
    L_{S} \approx 480~\text{m}\times \left(\frac{E_{s}}{100~\text{GeV}}\right) \left(\frac{6\times 10^{-4}}{g_{S}}\right)^2 \left(\frac{100~\text{MeV}}{m_S}\right)^4~.
\end{equation}
This makes far forward detectors like FASER well-suited for probing this model.

\section{Experimental Setup}\label{sec:detectors}

\begin{figure}[h]
    \centering
    \includegraphics[width=0.95\linewidth]{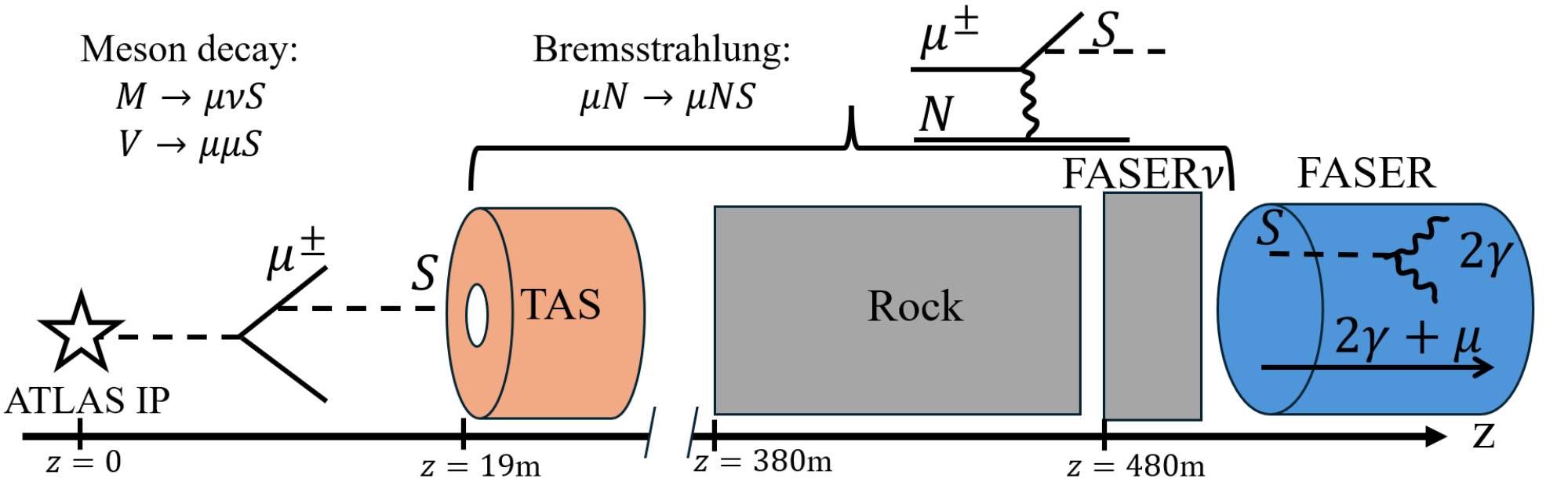}
    \caption{A simple cartoon illustrating the means of $S$ production through rare decays of pseudoscalar $(M)$ and vector $(V)$ mesons, 
    and muon Bremsstrahlung in the charged particle absorber (TAS), the rock, and in the FASER$\nu$ detector. We also show the two different signal types that we study: diphoton, and diphoton in association with the parent muon. For FASER2, the setup is similar, but with different sizes and locations of the various components, as described in the text.
    }
    \label{fig:cartoon}
\end{figure}

We now discuss the relevant components of the FASER experiment~\cite{FASER:2022hcn}, as well as some details of the LHC infrastructure. We first describe the FASER detector, and then work our way upstream with descriptions of the materials lying between FASER and the ATLAS IP which are relevant for $S$ production. An overview demonstrating the various production mechanisms is illustrated in ~\cref{fig:cartoon}. We will also discuss details of the FASER upgrade, FASER2, which is part of the Forward Physics Facility~\cite{Feng:2022inv}.

FASER is placed along the proton beam axis, 480~m downstream from the ATLAS IP while FASER2 is assumed to be 620~m away in the Forward Physics Facility cavern~\cite{Anchordoqui:2021ghd,Feng:2022inv,Adhikary:2024nlv}, also along the beam axis. The experiment is equipped with a front veto to reject incident muon events, a decay volume, a spectrometer for track momentum reconstruction, a preshower station for photon identification, and finally a calorimeter (see Fig.~1 of Ref.~\cite{FASER:2023tle}). We model the FASER(2) experiment as a cylindrical decay volume with a radius of 10 cm (1~m) and a length of 3.5~m (20~m), with the length including the spectrometer tracking volume as our signal does not require charged track reconstruction from the $S\rightarrow \gamma \gamma$ decay~\cite{Dienes:2023uve,Kling:2022ehv}. 

At the front of FASER, towards the ATLAS IP, is the FASER$\nu$ experiment, a tungsten emulsion detector which for our purposes will serve as a target for $S$ production via nuclear Bremsstrahlung, $\mu N \rightarrow \mu N S$. We model FASER$\nu$ (and its upgrade FASER$\nu$2~\cite{Feng:2022inv, SnowmassFASERnu2}) as a 1 (20) tonne target mass of tungsten plates interleaved with emulsion films that has transverse$\times$ longitudinal dimensions of $0.25\times0.3\times0.8~ {\rm m}^3$ ($0.4\times0.4\times6.6 ~{\rm m}^3$).

Between FASER$\nu$(2) and the primary LHC tunnel lies approximately 100 (240)~m of rock and concrete. Finally, 20~m from the IP in front of the quadrupole final-focusing magnets, lie the charged particle absorbers known as the TAS. During the HL-LHC era, the TAS will be upgraded to the TAXS to handle the increased particle flux. We model the TA(X)S as a 1.5~m long copper cylinder with a 25~cm radius; cut out from the cylinder is a small aperture for the beam pipe which has a radius of 1.7~cm (3~cm)\cite{Burkhardt:2120714,Aberle:2749422}. We explore both the rock and copper absorbers as an additional means of $S$ production.

Finally, we return to the FASER detector to discuss the preshower~\cite{FASER:2018bac} station in more detail as it is an important component for our study. Located at the back of the FASER detector just in front of the calorimeter, the preshower consists of two scintillator layers each preceded by a layer of tungsten. One of the main purposes of the preshower is to identify photon signals and separate them from neutrino interactions in the calorimeter~\cite{FASER:2018bac,Kling:2022ehv,FASER:2024bbl,Dienes:2023uve}. However, if a neutrino interacts close to or inside the preshower, this would be an irreducible background with the current experimental design~\cite{FASER:2024bbl} for photon energies relevant for our signal (${\cal O}(100)~{\rm GeV}$). Indeed, in the recent search for axion-like particles at FASER~\cite{FASER:2024bbl} which also had a multi-photon final state signature, neutrinos interacting in the preshower resembled signal events and thus required a rather large energy cut of 1.5~TeV to suppress them. To mitigate the background for photonic final states, the FASER collaboration is finishing the design of a high-granularity preshower upgrade~\cite{Boyd:2803084}. Simulations have shown that it can resolve two photons, each with at least 100 GeV energy, separated by $\gtrsim 200~\mu $m, with high ($\gtrsim 80\%)$ efficiency. The preshower upgrade is planned to be installed 
in time to capture the last $90~{\rm fb}^{-1}$~\cite{Boyd:2882503} of integrated luminosity of Run 3, although this amount is likely a conservative estimate as the upgraded preshower is now expected to capture more luminosity. 
For our study, 
we use the resolving power of the high-granularity preshower to mitigate backgrounds. We assume that the high-granularity preshower covers the entire radius of the spectrometer where photons are recorded. However, since the final impemented design may only cover a fraction of the spectrometer's transverse area, we explore the possibility of a reduced coverage in \cref{sec:appendix}, though we find that it does not significantly impact our results.

The FASER experiment has been approved for Run 4~\cite{Boyd:2882503}, and the Forward Physics Facility is in the planning stage. So we study two experimental scenarios: FASER during Run 4 and FASER2 during the HL-LHC era. For the former, we also include data that will be collected at the end of Run 3 with the expectation that the upgraded preshower will be installed by then. We assume FASER2 is equipped with a similarly capable preshower component. So, when presenting results, we will highlight discovery potential for FASER with the end of Run 3 + Run 4 luminosity ($90+680{\rm ~fb}^{-1}$) as well as with FASER2 during the full HL-LHC era ($3~{\rm ab}^{-1}$)

\section{Signal at FASER}\label{sec:signal}

In this section, we describe the signal at FASER from a muonphilic scalar, $S$, which decays to two photons. We begin by describing the two production modes of $S$, as each results in a different kinematic distribution of the signal. We also describe production of $S$ at various locations as they will probe different lifetime regions. We refer again to \cref{fig:cartoon} for an overview of production mechanisms and signal types.

The dominant production of $S$ proceeds through either rare three body decays of mesons or muon Bremsstrahlung in the intervening material between the ATLAS IP and FASER. Rare decays of mesons are a dominant production mechanism for a wide variety of BSM models~\cite{FASER:2018eoc,FASER:2023tle,FASER:2024bbl} at FASER. On the other hand, muon Bremsstrahlung of BSM particles at forward LHC experiments was first studied in Ref.~\cite{Ariga:2023fjg}, and subsequently in Ref.~\cite{Batell:2024cdl}. Additional production modes of $S$ exist, for example those involving its loop-suppressed photon interactions in neutral pseudoscalar meson decays or the Primakoff process in material near the ATLAS interaction point. These are expected to be subdominant and hence we do not consider them in this work.

For most of the production modes, the resulting signal in FASER will be two photons from $S$ decay, a commonly explored final state at forward detectors~\cite{FASER:2024bbl,Feng:2018pew,FASER:2018eoc,Feng:2022inv,Kling:2020mch,Kling:2022uzy}. Moreover, the present work is the first investigation of potential signal hiding in muon events; this corresponds to the case where the parent muon which produced the $S$ also enters the detector. The FASER experiment observes a muon flux of about 1 kHz~\cite{FASER:2021ljd}, corresponding to a few billion muons passing through FASER in Run 3, which are typically treated as a source of background. These events are currently discarded in neutrino and BSM searches, although some of them may contain BSM signatures, particularly when the BSM model is muonphilic. Although this signal is experimentally challenging because of the presence of the muon and the backgrounds it could induce, we make optimistic projections and analyze the signal distribution in the hope that the background can be controlled and these events can be studied.

\begin{figure*}
    \centering
    \includegraphics[width=0.45\textwidth]{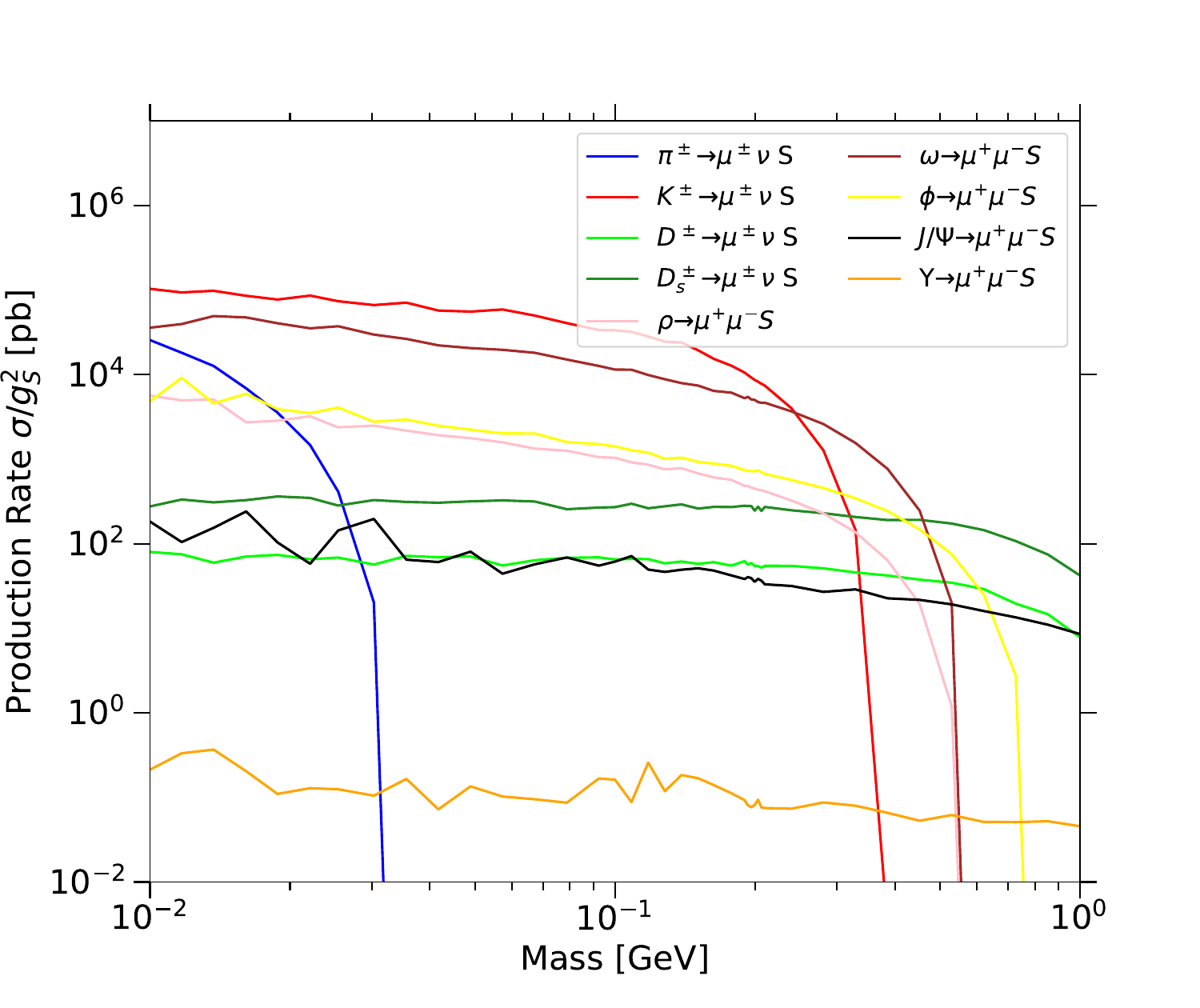} \hspace{0.5cm}
    \includegraphics[width=0.45\textwidth]{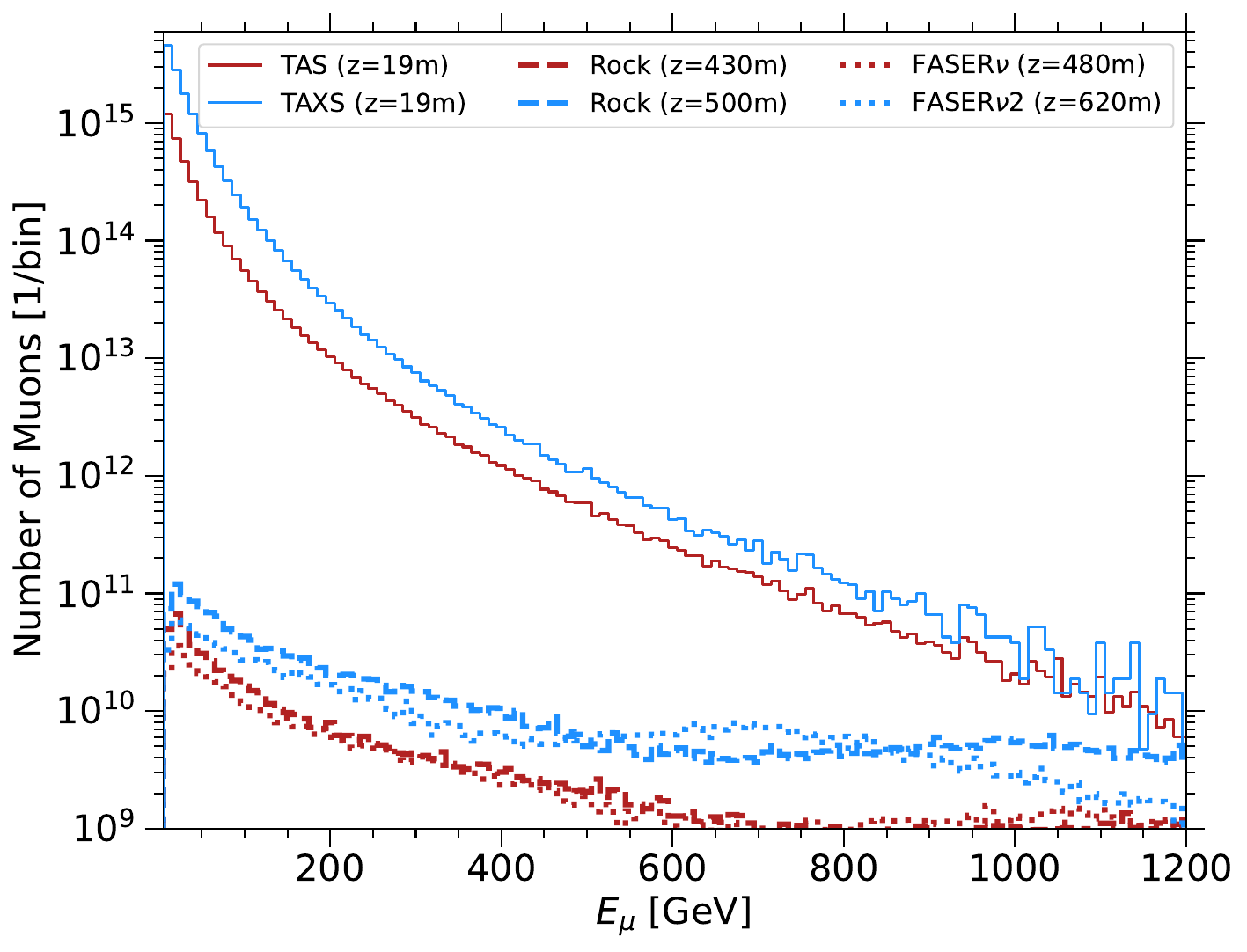}\\
    \caption{
    \textbf{Left:} Production rates of forward $S$ ($\theta < 0.2$ mrad) from rare three body decays of various mesons.
    Charged scalar mesons decay via a $W$ boson to a $\mu\nu S$ final state, and neutral vector mesons to a $\mu^{+} \mu^{-} S$ final state via a photon. For much of the mass range we are interested in, production from Kaons and $\omega$ mesons dominate.
    \textbf{Right:} 
    The number of muons passing through different materials relevant for Bremsstrahlung production of $S$. The red curves include the end of Run 3 + Run 4 luminosity (770 fb$^{-1}$) and the blue curves include the full high-luminosity (3 ab$^{-1}$). The charged particle absorbers, TA(X)S, are located 19.1 m from the ATLAS IP, and the FASER$\nu$(2) detector is 480 (620) m away. The intervening space consists of the LHC tunnel and rock, starting at 409 m. The midpoint between the start of rock and FASER$\nu$(2) detector is approximated as 430 (500)~m away from the ATLAS IP. 
    For further details of each target, see~\cref{sec:detectors} and \cref{sec:signal}.
    }
    \label{fig:fluxes}
\end{figure*}

\subsection{3 body Meson Decay - $2\gamma$}

The dominant means of producing $S$ is via rare decays of mesons. These mesons ($\pi,~K,~\omega,~\rho$, etc.) are copiously produced in $pp$ collisions at the ATLAS IP. Once produced, they can undergo rare three body decays, such as $K^{\pm}\rightarrow \mu^{\pm} \nu_{\mu} S$ for charged pseudoscalar mesons and $\omega\rightarrow \mu^{+} \mu^{-} S$ for neutral vector mesons. Ref.~\cite{Harris:2022vnx} explored a similar production mode, but only from kaons, whereas here we extend to include production from other pseudoscalar and vector mesons, which we find to be important. The muons are bent by the LHC magnets and also undergo multiple Coulomb scattering as they pass through $\sim \mathcal{O}(100)$~m of rock, changing their energy and direction. Hence, we safely assume that the muons produced from these rare meson decays miss the FASER detector, unlike $S$ which travels forward unimpeded. Indeed this assumption can be further justified by the comparing the relatively smaller muon flux at FASER$\nu$ with that at TAS in the right panel of~\cref{fig:fluxes}. At 100 GeV (1 TeV) the muon flux at FASER$\nu$ is smaller than the flux at TAS by a factor of 1000 (100). So in this case we expect only $S$ to enter the detector (and for pseudoscalar meson decay, an accompanying neutrino which is extremely unlikely to interact).

We model the production of $S$ via meson decay in \texttt{FORESEE}~\cite{Kling:2021fwx}, which includes the SM meson spectrum in the forward direction. It is worth noting that some pseudoscalar mesons are long-lived, and thus are more likely to interact somewhere in the LHC before decaying. We only study $S$ production from those which decay before the TAS---it is possible that there may be some additional reach afforded by propagating them through the magnetic fields and simulating their decay, but we do not consider this here.
The differential branching ratio for a charged pseudoscalar $M^{\pm}$ to decay into $\mu^{\pm}\nu_{\mu} S$ and neutral vector meson $V$ to decay into $\mu^{+}\mu^{-} S$
is given by~\cite{Carlson:2012pc,Mitra:2021uun}
\begin{widetext}
\begin{equation}
\label{eq:3body-scalar}
\begin{aligned}
&\frac{d^2\text{BR} (M^{\pm} \to \mu^{\pm} \nu_{\mu} S) }{dE_S dQ^2}
=\frac{  m_M g_S^2 \times  \text{BR} (M^{\pm} \to \mu^{\pm} \nu_{\mu})  }{ 8 \pi^2 m_\mu^2 (m_M^2-m_\mu^2)^2 (Q^2-m_\mu^2)^{2} }   \\
&\times \Big( ( m_M^2 - 2 m_M E_S + Q^2 ) Q^2 (Q^2-m_\mu^2) - (Q^4-m_\mu^2 m_M^2)(Q^2+m_\mu^2-m_S^2) + 2 m_\mu^2 Q^2 (m_M^2-Q^2) \Big) ~,
\\
\\
&\frac{d^2\text{BR} (V \to \mu^{+} \mu^{-} S) }{dE_S dQ^2}
=\frac{1}{\Gamma_V}\frac{  2\alpha^2 g_S^2 f_V^2  }{ 27 \pi m_V^4 \left[(Q^2-m_{\mu}^2)  (Q^2-2E_Sm_V-m_{\mu}^2)\right]^2}   \\
&\times \Big( \left[(Q^2-m_{\mu}^2)^2 + (Q^2-2E_Sm_V-m_{\mu}^2)^2\right](4m_{\mu}^2-m_S^2)(2m_{\mu}^2+m_V^2) \\
&- 4E_Sm_V(4m_{\mu}^2-m_S^2+E_Sm_V)(Q^2-m_{\mu}^2)(Q^2-2E_Sm_V-m_{\mu}^2) \\
&-2(Q^2-m_{\mu}^2)(Q^2-2E_Sm_V-m_{\mu}^2)(8m_{\mu}^4+m_S^4+m_{\mu}^2(4m_V^2-6m_S^2))\Big)~.
\end{aligned}
\end{equation}
\end{widetext}
Here $\alpha$ is the fine structure constant, $m_M$ and $m_{\mu}$ are the mass of the pseudoscalar meson and muon respectively, $E_S$ is the energy of $S$, $Q$ is the invariant mass of the $(\mu^{\pm},~S)$ system, and $\text{BR} (M^{\pm} \to \mu^{\pm} \nu_{\mu})$ is the branching ratio for the pseudoscalar meson to decay into a muon and a neutrino.
$m_V,~f_V$ and $\Gamma_V$ are the mass, decay constant and total decay width of the vector meson, $Q$ is again defined as the invariant mass of the ($\mu,~S$) system with $\mu$ being the muon from which $S$ originates.
In the left panel of \cref{fig:fluxes}, we show the production rates for $S$ from various mesons. For the masses in which we are interested, production from $K$ and $\omega$ dominate in roughly equal amounts, demonstrating the importance of including production modes other than $K$, in particular the $\omega$ meson.

Once produced, only the $S$ particles with sufficiently small transverse momenta travel towards FASER uninterrupted. As is shown in left panel of \cref{fig:fluxes}, we only consider $S$ particles that are produced with an angle less than 0.2 mrad from the beam axis for FASER.
We simulate the propagation to and subsequent decay of $S$ particles inside FASER using \texttt{FORESEE}~\cite{Kling:2021fwx}.
Since they are produced close to the ATLAS IP they have to travel a distance of $\sim 500$~m to reach the FASER detector. In the left panel of \cref{fig:kinematics}, we show the energy spectrum of the diphoton system from $S$ decay inside FASER for two benchmark points: $(m_S,g_S)$ = (35 MeV, $2.5\times 10^{-3}$), and (100 MeV, $6.3\times 10^{-4}$).
In our analysis we also make use of the high granularity preshower which can identify two sufficiently separated photons, and so in the middle panel of \cref{fig:kinematics}, we show the distribution of the two photon's separation at the preshower location, defined as $\Delta_{\gamma\gamma}$. An energetic diphoton pair has a smaller separation, as the separation goes as $\Delta_{\gamma\gamma}\sim m_S/E_S$. Studies by the FASER collaboration indicate that the 2 photons can be resolved with high efficiency~\cite{Boyd:2803084} for $E_{\gamma\gamma}>200 $~GeV, and $\Delta_{\gamma \gamma}>200~\mu$m. To mitigate backgrounds, we apply these cuts on our signal and find that there still exists a significant event rate.

\subsection{Muon Bremsstrahlung}

Muons are capable of penetrating the rock and LHC infrastructure that lie between the ATLAS IP and FASER. 
While traversing this distance, the muons pass through a variety of targets where they can produce new particles, see~\cref{fig:cartoon}. In what follows we will discuss the contribution of $S$ production and signal via Bremsstrahlung in the charged-particle absorbers (TA(X)S), the rock in front of FASER, and the tungsten target in FASER$\nu$.\footnote{There are additional opportunities for Bremsstrahlung production of $S$, e.g. in the TAN neutral particle absorbers, or in the material surrounding the beam pipe, though we do not model these productions here.} For all targets, we simulate the $\mu N\rightarrow \mu N S$ process in \texttt{MadGraph}~\cite{Alwall:2011uj} using the appropriate form factor~\cite{Jodlowski:2019ycu} for each nuclear target\changed{}{, for a nucleus $N$}. The number of signal events is then given by integrating over the incident muon flux, and the differential cross-section for $S$ production and the decay probability:
\begin{equation}
\label{eq:brem}
\begin{split}
        N_{\rm Signal} &= n_{\rm target}\int dE_{\mu_i}d\Omega_{\mu_i}dE_Sd\Omega_S dz\times \\ 
        &\left(\frac{dN_{\mu_i}}{dE_{\mu_i}  d\Omega_{\mu_i}}\right) \left(\frac{d\sigma (E_{\mu_i})}{d{\Omega_S}dE_S}\right) P_{\rm decay}(E_S,z)~.
\end{split}
\end{equation}

Here, $n_{\rm target}$ is the atomic number density of the target, and the first term in parentheses is the muon flux passing through the target 
as a function of the incident muons' energy and angular distribution ($E_{\mu_i},\Omega_{\mu_i}$).
The second term in parentheses is the differential cross section for $S$ production via muon Bremsstrahlung.
Here $E_S,\Omega_S$ describe the produced scalar's energy and angular distribution,
$z$ is the longitudinal coordinate parallel to the beam axis and integrates up to the target depth, and $P_{\rm decay}(E_S,z)$ is the probability for $S$ with energy $E_S$ to decay at a distance $z$. Here, we compute~\cref{eq:brem} via a Monte Carlo simulation.

\subsubsection{TAS - $2\gamma$}
As discussed in \cref{sec:detectors}, the TAS and TAXS are the copper charged-particle absorbers near the ATLAS IP during Run 3 and HL-LHC era respectively, which will experience an incident muon flux before the magnets have influenced their trajectory (see Fig.~1 of Ref.~\cite{FASER:2024ykc}). In principle, $S$ could be produced in the other materials surrounding the absorbers, but these lie at large radii from the beam axis and thus have a subdominant contribution to the signal at FASER because $S$ is produced with small angular deviations relative to the incident muon. 

We simulate the incident muon flux by generating inelastic proton collisions as predicted by \texttt{Pythia 8}~\cite{Sjostrand:2007gs}, sampling the hadron decay up to the absorbers' location at $z=19.1$~m, with the muon flux being dominated by pion, kaon, and charm hadron decays. We note that there is some uncertainty in the forward hadron flux~\cite{Fieg:2023kld,FASER:2024ykc} which we do not consider in our analysis here, though it is expected to not significantly impact our results. We show the resulting number of muons passing through the absorbers in the right panel of \cref{fig:fluxes} including their respective luminosities. The total number of muons passing through the absorbers is $\gtrsim 10^4$ times larger than the that passing through the FASER detector~\cite{FASER:2021mtu}, owing primarily to the deflection of the muons by the magnets and subsequent multiple Coloumb scattering in the intervening material.

Because muons will be affected by both the magnets and the intervening material, it is unlikely that a muon which produces an $S$ will also intercept the FASER detector. Thus, the diphoton signal from this production mode is identical to the contribution to the signal from rare meson decays.  However, given the different production mechanisms, the signal kinematics are different. 

\begin{figure*}[th!]
    \centering
    \includegraphics[width=0.32\textwidth]{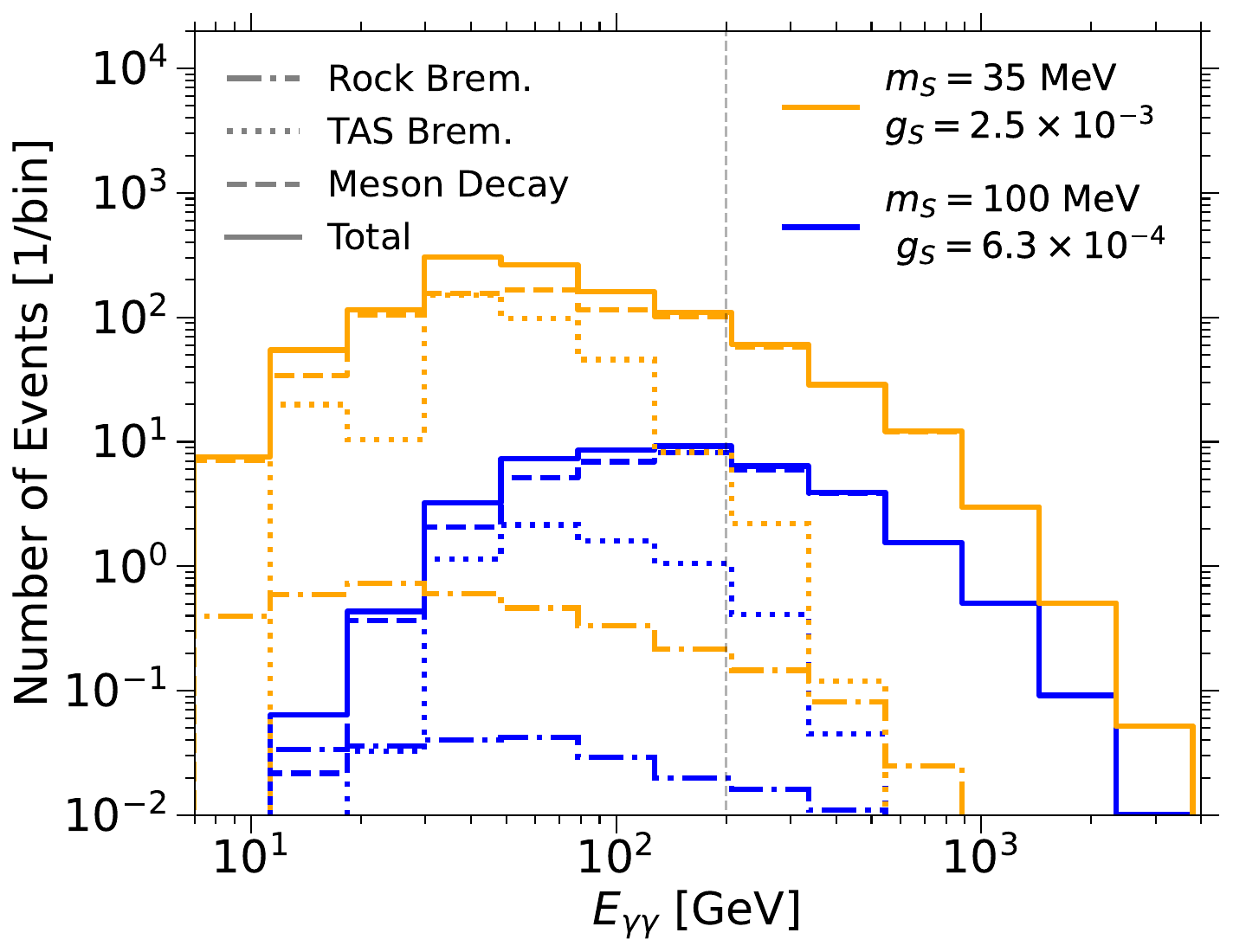}
    \includegraphics[width=0.32\textwidth]{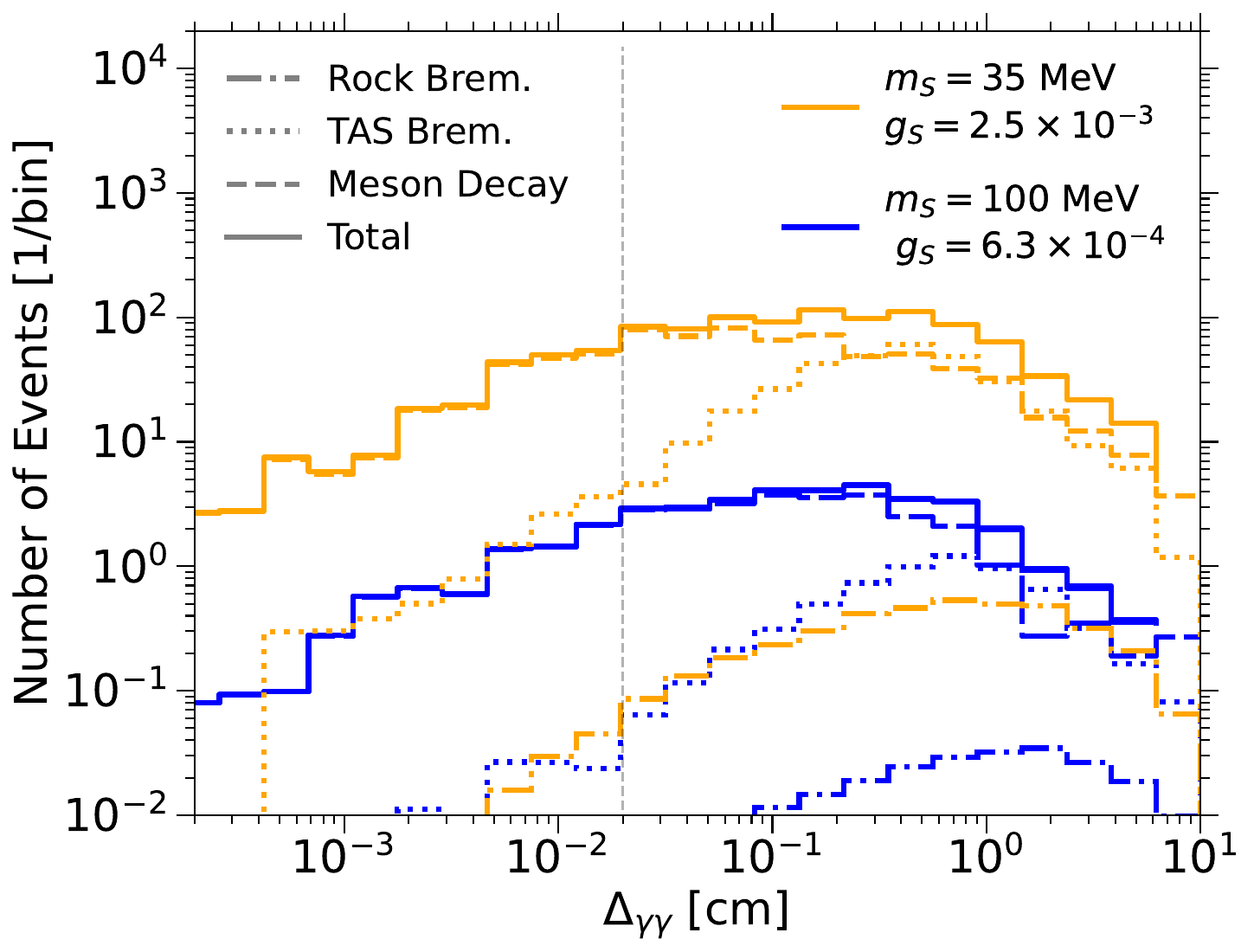}
    \includegraphics[width=0.32\textwidth]{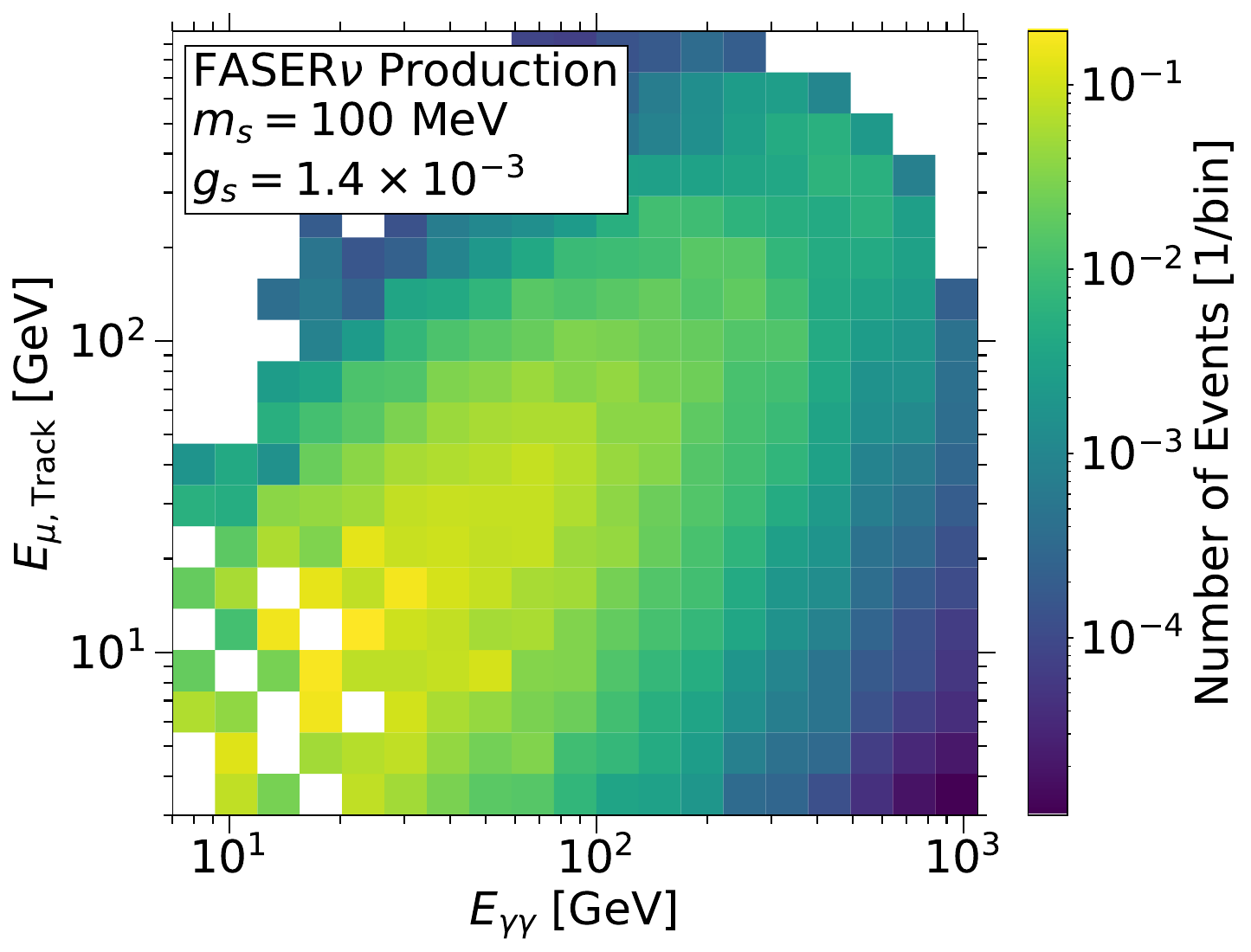}
    \caption{Energy $E_{\gamma\gamma}$ (left), and separation spectra $\Delta_{\gamma\gamma}$ (middle), of the diphoton signal coming from decays of $S$ for 770 fb$^{-1}$. We show it for two benchmark points: $(m_S,g_S)$ = (35 MeV, $2.5\times 10^{-3}$) in orange, and (100 MeV, $6.3\times 10^{-4}$) in blue.
    The spectra are also shown for the three production modes: rare meson decays (dashed), Bremsstrahlung in the TAS (dotted), and Bremsstrahlung in the rock (dashdotted). $S$ produced via rare meson decays are more energetic than $S$ from other production modes resulting in a harder diphoton energy spectra, and a smaller photon separation ($\Delta_{\gamma\gamma}\sim m_S/E_S$). In the vertical gray dashed lines, we also show the kinematic cuts we employ to eliminate background, $E_{\gamma \gamma}>200$ GeV and $\Delta_{\gamma \gamma}>200~\mu$m. In the rightmost panel, we show the signal distribution for $S$ production in FASER$\nu$, consisting of two photons and the parent muon, for $(m_S,g_S)$=(100 MeV, $1.4\times 10^{-3}$). On the vertical axis, we show the energy of the final-state muon which can be reconstructed by the trackers, and on the horizontal axis we show the summed energy of the two photons and the typical energy deposited by the muon in the calorimeter, with the two photons' energy dominating. 
    For this panel, we require $\Delta_{\gamma \gamma}>200~\mu $m.
    } 
    \label{fig:kinematics}
\end{figure*}

We simulate $S$ production in the copper ($Z=29$) absorbers, their propagation, and decay in FASER. In~\cref{fig:kinematics}, we show the summed energy spectra of the two photons ($E_{\gamma \gamma}$, left panel) and their separation ($\Delta_{\gamma \gamma}$, middle panel) at the preshower detector for the same benchmark points as before.
We find that the requirements we impose on $E_{\gamma \gamma}$ and $\Delta_{\gamma \gamma}$ render this production mode subdominant to $S$ production from meson decay at these benchmark points, though we still consider production in the TAS in the full analysis. We find that if the requirement of $E_{\gamma \gamma}>$200~GeV can be relaxed, production in the TAS alone is sensitive to unprobed parameter space, though it is still subdominant to $S$ from meson decay.
We also note that the signal from TAS production has relatively lower energy signal photons and thus larger diphoton separation as compared to the signal from meson decay. This is partly due to the fact that the $S$ particles produced in the copper absorbers have comparatively lower energy, as the highest energy muons pass through the aperture in the middle and do not encounter the copper at all.

 \subsubsection{Rock - $2\gamma$}
To simulate $S$ production in the rock ($Z=11$~\cite{ParticleDataGroup:2020ssz}) near FASER we must obtain the muon flux after it has passed through the magnets and intervening materials. The CERN sources, targets, and interactions group performed a detailed FLUKA~\cite{Ferrari:2005zk,Battistoni:2015epi} simulation to obtain the muon flux through a $1\times1$ m$^2$ area, 409~m from the ATLAS IP, which is the approximate location where the collision axis exits the concrete lining of the LHC tunnel and enters the exterior rock (see Fig.~2 of Ref.~\cite{FASER:2021mtu}). 
As an approximation, we use the muon flux at the midpoint of the rock for production of $S$.
To this end we propagate the muons from 409~m to $430 ~(500)$~m for FASER(2), taking into account the multiple Coulomb scattering in the rock.
In the right panel of \cref{fig:fluxes}, we show the number of muons passing through this location within a transverse area defined by a circular cross-section with a radius matching that of the cylindrical FASER(2) decay volume.

We then sample the spatial and momentum distributions of the muons for $S$ production, analogous to production in the TAS. For production in the rock, however, we must ensure that the outgoing muon does not pass through FASER. For this, we propagate the muons all the way to FASER, and apply a muon energy dependent efficiency factor to the signal rate that accounts for the probability that the muon scatters away (this factor is found to be about $\approx$50\%). Hence, the signal we see from this production mode is also two photons, identical to the meson decay and Bremsstrahlung in the TAS. 

\subsubsection{FASER$\nu$ - $\mu+2\gamma$}
Given the dense FASER$\nu$ target (tungsten $Z=74$), it could serve as an additional means of $S$ production via Bremsstrahlung. 
However, in this case the parent muon will trigger the FASER's front veto, and is very likely to enter the FASER detector volume, and the signal will be two separated photons along with the parent muon. 
 
The advantage of this production mechanism over the others is that the photons from $S$ are more likely to intercept the spectrometer. Moreover the parent muon will also enter the detector, whose momentum can be reconstructed, thus offering an additional handle on event reconstruction. However, it is expected that this signal will prove to be an experimental challenge due to additional muon interactions in the detector, including those that may mimic our signal.\footnote{This signal can be supplemented by considering $S$ production in the rock and requiring that the muon also intercept FASER. As the backgrounds and systematics for this signal are not well-understood, we do not consider this additional contribution and instead only consider production in FASER$\nu$ to demonstrate the physics case.} 

Some of the observables which can be used to classify this signal are:
the front veto is triggered, one muon track with reconstructed momentum, two separated photons in the preshower, and the energy deposited in the calorimeter.
We perform a simple analysis, placing cuts only on the diphoton energy and separation, assuming the relatively unstudied backgrounds can be controlled.  
We emphasize that this analysis is very optimistic. Nevertheless, we wish to assess the full opportunity that this production mechanism can offer, to lay the groundwork for future analyses with dedicated detector simulations.

After propagating the muons to FASER$\nu$, we simulate the diphoton signal distribution, similar to the simulations for $S$ production discussed previously. The key difference is the measurement of the final-state muon which emerges from FASER$\nu$. 
The energy of this observed muon is correlated with the energy of $S$ and thus the diphoton's energy. The energy measured in the calorimeter, is the sum of the two photon energies and the energy that the muon deposits. There is not yet any study of the preshower's ability to resolve two photons when a muon is present, so for now we optimistically assume that the two photons can still be effectively resolved after the same required kinematic cuts are imposed. When presenting the signal sensitivity from this production mechanism, we do not impose any cuts on the measured muon momentum.

In the right panel of \cref{fig:kinematics} we show the signal event distribution of the muon's energy after $S$ production, and the energy deposited in the calorimeter for $\Delta_{\gamma \gamma}>200~\mu{\rm m}$
for a benchmark point of $(m_S, g_S)$ = (100 MeV, 1.4 $\times 10^{-3}$). To estimate the energy measured by the calorimeter, we sum the signal photons' energy, and add the typical energy deposited in the calorimeter by the accompanying muon; the former dominates, with the muon typically depositing $\lesssim 500~{\rm MeV}$~\cite{ParticleDataGroup:2020ssz}. The muon energy reconstructed by the tracker is correlated with the photons' energy measured in the calorimeter, and the spread in their relation is due to the $S$ energy distribution in the $\mu N \rightarrow \mu N S$ production process.

\subsection{Background}
\label{sec:background}
We now discuss some details concerning background for the diphoton signal, and the muon + diphoton signal separately.
For the diphoton signal, its background at FASER has been well-studied in the context of axion-like particle searches~\cite{FASER:2024bbl} where the final state consists of multiple photons with no associated charged tracks. 
However, it is important to note that, with the current preshower detector, these photons are not resolved individually, but will instead be observed as a total energy deposition in the calorimeter.

The dominant background for multi-photon final states comes from neutrinos interacting in the detector, typically producing a number of charged particles through deep-inelastic scattering---these charged particles can potentially be used as a veto using the tracking stations.
For the axion-like particle search, it was estimated that 19 neutrinos would interact in the preshower itself and would pass preliminary cuts to resemble the diphoton signature for an integrated luminosity of 57.7 fb$^{-1}$. This expected background was further reduced to $<$ 1 event by requiring large energy depositions of $E>1.5$~TeV in the calorimeter. In the absence of any other handles on the background, this energy cut would significantly weaken our reach (see \cref{sec:appendix}).
For our projections, we take advantage of the planned high-granularity preshower upgrade at FASER, which initial studies have shown can resolve two 100 GeV photons with separation $\Delta_{\gamma \gamma}\gtrsim 200~\mu$m. We assume that the high-granularity preshower, in combination with the tracking spectrometer and calorimeter, can completely eliminate the neutrino background.

For $S$ production in FASER$\nu$, the signal is the final-state muon and the two photons. To our knowledge this is the first attempt to pursue a BSM signature in the many muons events that are regularly observed at FASER, and thus the backgrounds are not well-understood. The muons will typically leave relatively small, $\lesssim 100~{\rm MeV}$, energy deposits, but due to the enormous flux we also expect a large number of harder interactions which produce a shower of secondary particles~\cite{FASER:2023tle}. In particular, a muon can interact somewhere in the detector which may result in two photons, for example through a Bremsstrahlung interaction or through $\pi^0$ production. These photons would typically have a much smaller energy than those of our signal, but it is possible that more rare, harder interactions amount to an important background.
While we provide some preliminary considerations, this unexplored signal requires a detailed detector simulation.
We show the most optimistic, background free projections, under the assumption that the backgrounds can be controlled in this experimentally challenging channel.

\section{Results}\label{sec:results}

\begin{figure*}[th!]
    \centering
    \includegraphics[width=0.48\textwidth]{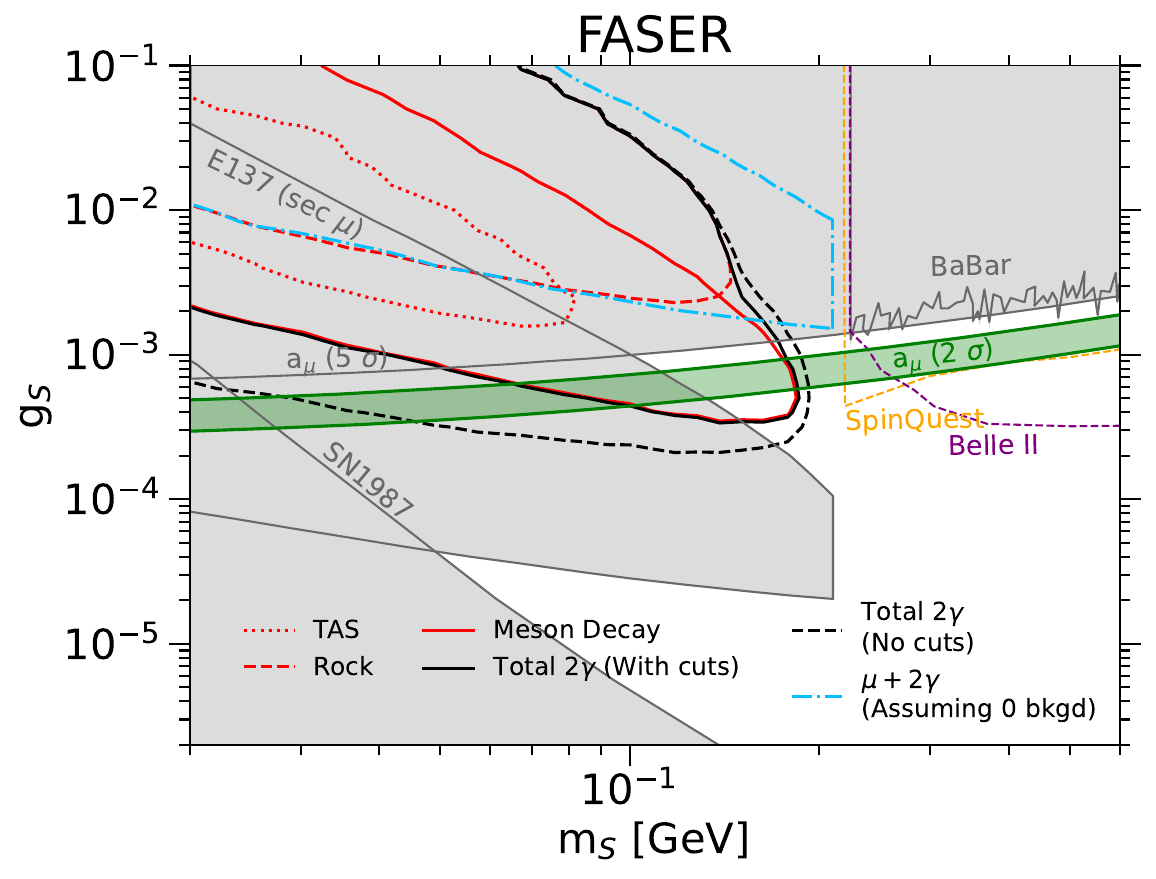} 
    \includegraphics[width=0.48\textwidth]{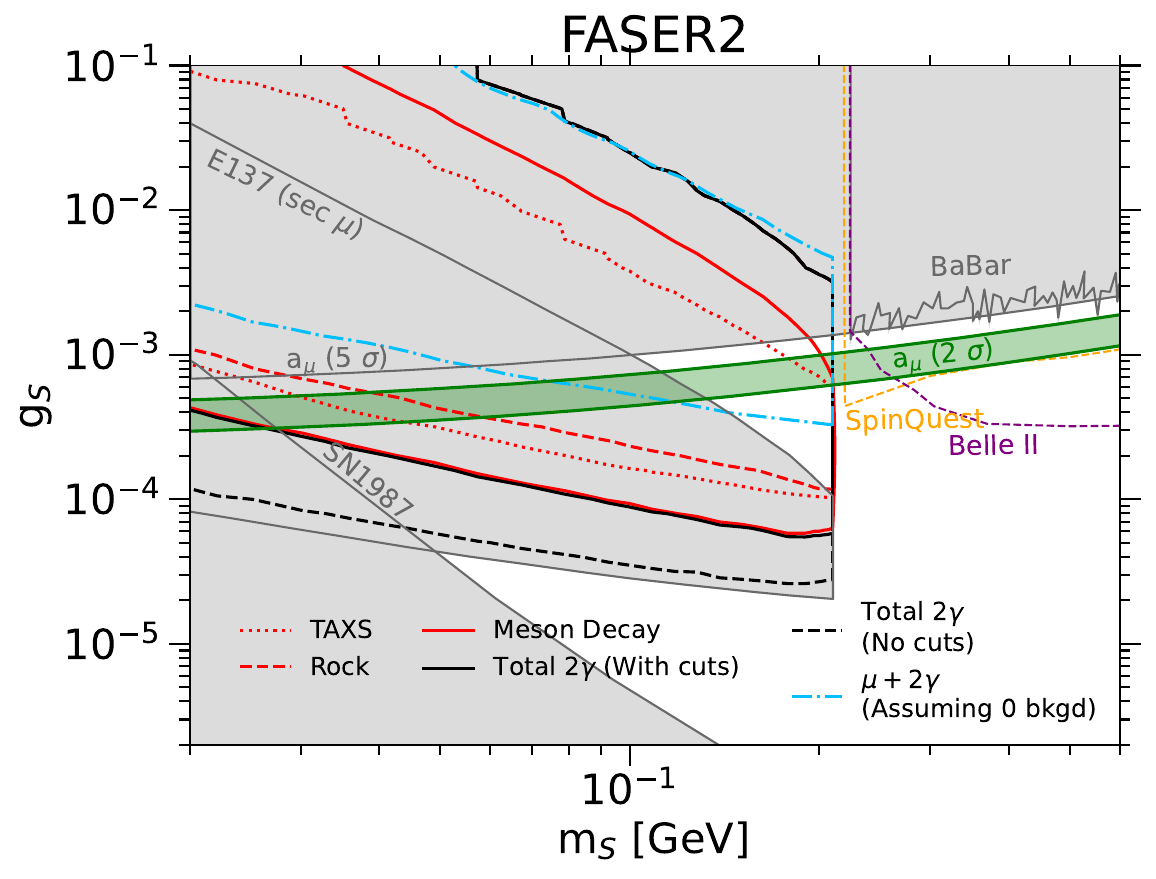} 
    \caption{Projected exclusion at 90\% C.L. for FASER (left) using the end of Run 3 + Run 4 luminosity (770 fb$^{-1}$) and FASER2 (right) during the HL-LHC (3 ab$^{-1}$). 
    For the diphoton signal, a background free search requires the two photons in the final state to be spatially resolved. For the upgraded preshower this means $E_{\gamma\gamma}>200$ GeV and $\Delta_{\gamma\gamma}>200~\mu$m.
    In the solid black line, we show the 3 event contour line after placing these cuts.
    We also show the limits without placing the cuts in the dashed black line, to show the best case scenario where the requirements on $E_{\gamma \gamma}$ and $\Delta_{\gamma \gamma}$ are completely removed. This also demonstrates that the cuts primarily affect the reach for relatively smaller $g_s$.
    The red lines show the contribution from different production modes after imposing the cuts: rare meson decays (solid red), Bremsstrahlung production in TA(X)S (dotted red) and in the rock (dashed red). 
    The qualitatively different signal coming from muon Bremmstrahlung inside FASER$\nu$ is shown separately in dash-dotted light-blue after placing the same cuts, and assuming no backgrounds.
    The green band shows the preferred parameter space to solve the $(g-2)_{\mu}$ anomaly. Existing constraints are shown in gray and come from measurement of $a_{\mu}$~\cite{Muong-2:2021ojo,Muong-2:2023cdq}, E137~\cite{Bjorken:1988as,Marsicano:2018vin}, SN1987~\cite{Croon:2020lrf,Rella:2022len}, and BaBar~\cite{BaBar:2016sci,Capdevilla:2021kcf,Krnjaic:2019rsv,Chen:2017awl}. Projections from ongoing experiments such as SpinQuest~\cite{Apyan:2022tsd,Forbes:2022bvo} (orange dashed) and Belle II~\cite{Capdevilla:2021kcf} (purple dashed) are also shown which can cover the $(g-2)_{\mu}$ preferred band for heavier masses.
    } 
    \label{fig:reach}
\end{figure*}

In \cref{fig:reach}, we show the projected reach at 90\% C.L in the $m_S,g_S$ plane for the FASER experiment (left panel) using the end of Run~3 + Run~4 data (770~fb$^{-1}$) and the FASER2 experiment (right panel) for the entire HL-LHC era (3~ab$^{-1}$). 
In solid black, we present exclusion contours requiring 3 diphoton signal events after applying cuts on photon energy $E_{\gamma \gamma}>$200~GeV, and separation $\Delta_{\gamma \gamma}>200~\mu$m at the preshower station.  
To illustrate the most optimistic scenario, we also show our reach without requiring cuts in the dashed black line. 

The photon separation can also be thought of as an upper bound on the energy as  $\Delta_{\gamma \gamma}\sim m_S / E_S$ where $E_S$ is the energy of the produced $S$ and is proportional to the photons' energy. So, in effect, the requirement of $\Delta_{\gamma \gamma}>200~\mu$m  mostly cuts sensitivity on the short lifetime regime along the upper contour; on the other hand the $E_{\gamma \gamma}>$ 200 GeV requirement cuts sensitivity on the long lifetime regime on the lower contour.  However, in the short lifetime regime the event rate falls exponentially with $g_S$ and so the photon separation turns out to not have a large impact on the exclusion reach, though it is important for control of the neutrino background.

To illustrate the relative production mechanisms' contribution to the total reach we also show in red the 3 event contour lines for the diphoton signal for each production mode separately after requiring the kinematic cuts, with rare meson decays shown as the solid line, production in the TA(X)S as the dotted line, and production in the rock as the dashed line.
For $S$ production in FASER$\nu$, the signal is qualitatively different due to the presence of the parent muon, and so we show it in the dash-dotted light-blue colored line after applying the same cuts as above. 
This contribution is not included in the summed contours (solid and dashed black lines). It is important to note that 
despite the kinematic cuts,
this may be an experimentally challenging signal and is subject to backgrounds that are not studied in detail here. 
At FASER, we find that it only contributes additional reach at the largest couplings, which are already excluded; at FASER2 however, there is a significant event rate in unprobed parameter space from production in FASER$\nu$2. However, it is worth noting that the signal arising from $S$ production in FASER$\nu$2 can only probe parameter space that can also be probed by the diphoton signature at FASER2.

We find that $S$ production via rare meson decay contributes the dominant event rate at FASER for unprobed parameter space around $m_S$=150 MeV and $g_S=5\times10^{-4}$, with $S$ production in the TAS and rock providing subdominant contributions.
In contrast to the other production modes, $S$ production in the rock and FASER$\nu$ is able to probe larger couplings due to the relatively shorter decay length.
On the other hand, at FASER2, the much larger decay volume and luminosity lead to significantly increased event rates for all production mechanisms, and the exclusion contour for each saturates the $m_S<2m_{\mu}$ limit which is required for the diphoton signal. 
Moreover, while we are generally interested in searching for muonphilic scalars, we show the part of parameter space that can solve the $(g-2)_{\mu}$ anomaly in the green shaded band. 
We find that FASER2 is able to completely cover the open parameter space around 150 MeV which resolves this anomaly.

In the gray shaded regions, we show parts of parameter space that are already excluded by previous experiments and searches.
Existing constraints include those coming from the measurement of $a_{\mu}$~\cite{Muong-2:2021ojo,Muong-2:2023cdq}, E137~\cite{Bjorken:1988as,Marsicano:2018vin}, SN1987~\cite{Croon:2020lrf,Rella:2022len}, 
and BaBar~\cite{BaBar:2016sci,Capdevilla:2021kcf,Krnjaic:2019rsv,Chen:2017awl}.
Some ongoing experiments like SpinQuest~\cite{Apyan:2022tsd,Forbes:2022bvo} (orange dashed) and Belle II~\cite{Capdevilla:2021kcf} (purple dashed) which can close the $(g-2)_{\mu}$ preferred region for heavier masses are also shown.
For other proposed experiments and analysis studying similar models, see Refs.~\cite{Batell:2016ove,Cesarotti:2023udo,Blinov:2024gcw,Kahn:2018cqs}

\section{Conclusion}\label{sec:conclusion}
In this work, we have investigated the potential to discover long-lived muonphilic scalars which decay to two photons at the FASER experiment and its proposed upgrade FASER2. Muonphilic scalars are particularly interesting as they may be the reason for the discrepancy between the experimental measurement and the theoretical prediction of the magnetic moment of the muon. They may also serve as mediators to a dark sector to achieve the observed dark matter relic abundance. 

We have calculated the scalar production near the ATLAS IP via rare meson decays, as well as production in the LHC infrastructure, the rock near FASER(2), and within the FASER$\nu$(2) experiment itself. In order to mitigate background, we make use of the planned preshower upgrade~\cite{Boyd:2803084}, which is expected to be installed before the end of Run 3 and is capable of resolving photons separated by $\gtrsim 200~\mu$m.
We find that these forward detectors are sensitive to previously unexplored regions of parameter space, underscoring the unique capability of the forward region of the LHC, effectively functioning as a TeV-scale muon beam dump. We note that our projections are likely a conservative estimate as there are additional production modes we do not consider such as loop-suppressed processes, or muon Bremsstrahlung in other materials within the LHC infrastructure.

To the best of our knowledge, for the first time, we assess the potential for the billions of muons that are regularly observed by FASER to be associated with new physics (complementing other proposed searches at FASER$\nu$(2)~\cite{Ariga:2023fjg,Batell:2024cdl}).
These events have not been considered for any new physics studies and are currently discarded as background events to other neutrino and BSM searches. We note, however, that using muon events for BSM searches may be experimentally challenging due to the large muon-induced background. We nevertheless study the kinematic features and present 
the optimistic sensitivity in the hope that the background can be controlled in future studies. 

Using only the diphoton events, we find that FASER is sufficiently sensitive to mostly cover a gap in the muonphilic scalar parameter space that can solve the $(g-2)_{\mu}$ anomaly for $m_S<2m_{\mu},$ and that it's upgrade, FASER2, will be able to completely cover this gap. This work highlights the underappreciated potential of the world's highest-energy muon beam and presents the need for further exploration to fully harness the physics opportunities offered by the LHC in the forward direction.

\section*{Acknowledgements}
We would like to acknowledge Akitaka Ariga, Brian Batell, Jamie Boyd, Jonathan L.~Feng, Tomohiro Inada, Felix Kling, Toni M\"akel\"a, Thong Nguyen, Ken Ohashi, and Sebastian Trojanowski for useful discussions. We also thank Sudip Jana and Vishnu P.~K. for their helpful suggestions during the early stages of this work.
We also thank Felix Kling for sharing code used for the muon propagation in matter. 
The work of R.M.A and M.F. is supported in part by U.S.~National Science Foundation Grants PHY-2111427 and PHY-2210283. The work of R.M.A was further supported by the Simons Investigator Award \#376204. M.F. was further supported by NSF Graduate Research Fellowship Award No. DGE-1839285.

\appendix
\section{Impact of Energy Cuts, Luminosities, and Preshower Coverage on Exclusion Reach}\label{sec:appendix}

\begin{figure*} \centering \includegraphics[width=0.48\textwidth]{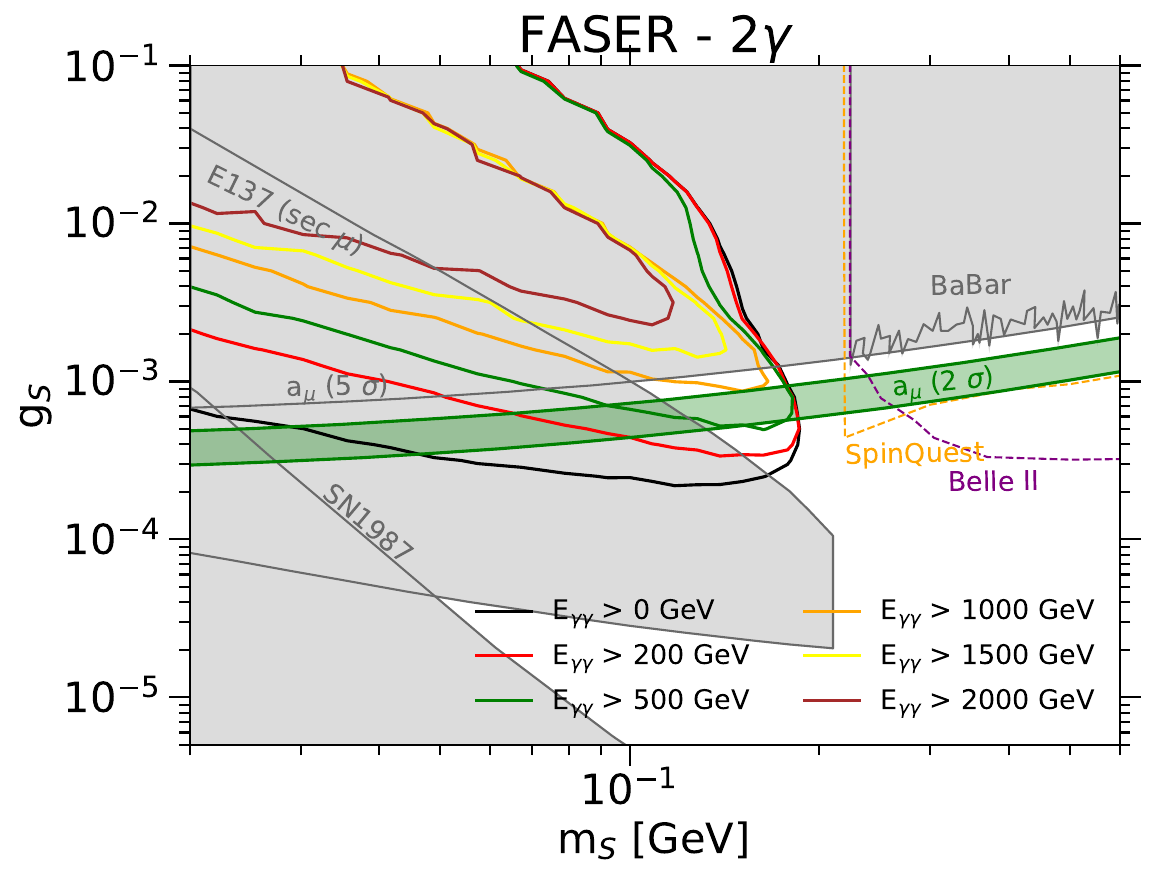} \includegraphics[width=0.48\textwidth]{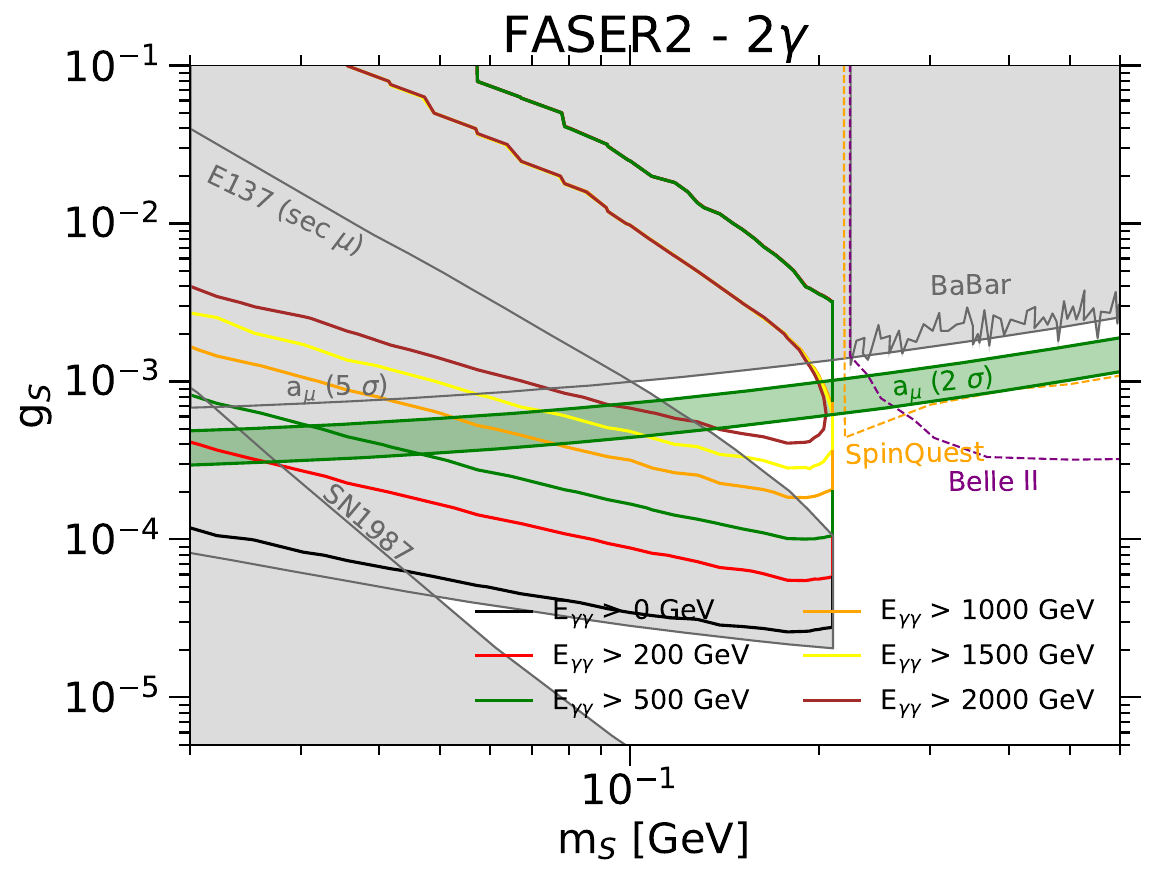} \includegraphics[width=0.48\textwidth]{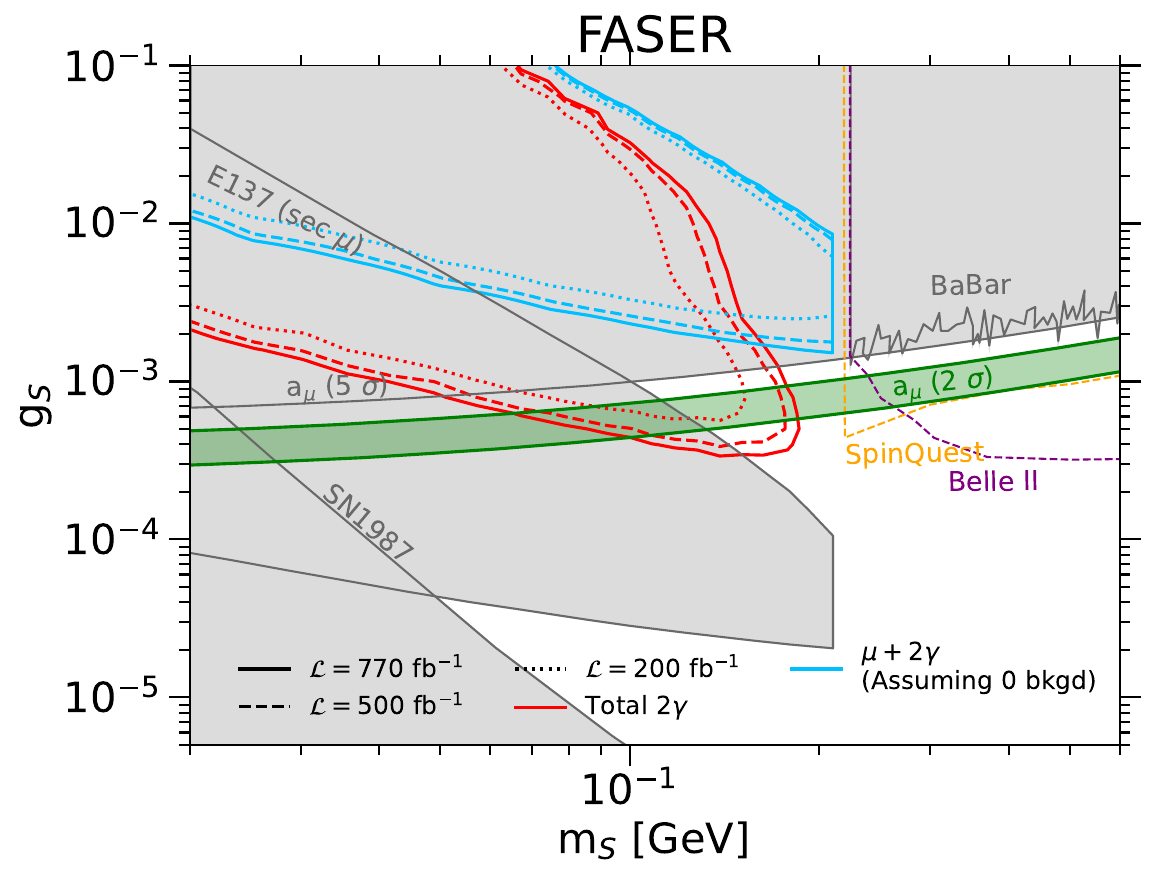} \includegraphics[width=0.48\textwidth]{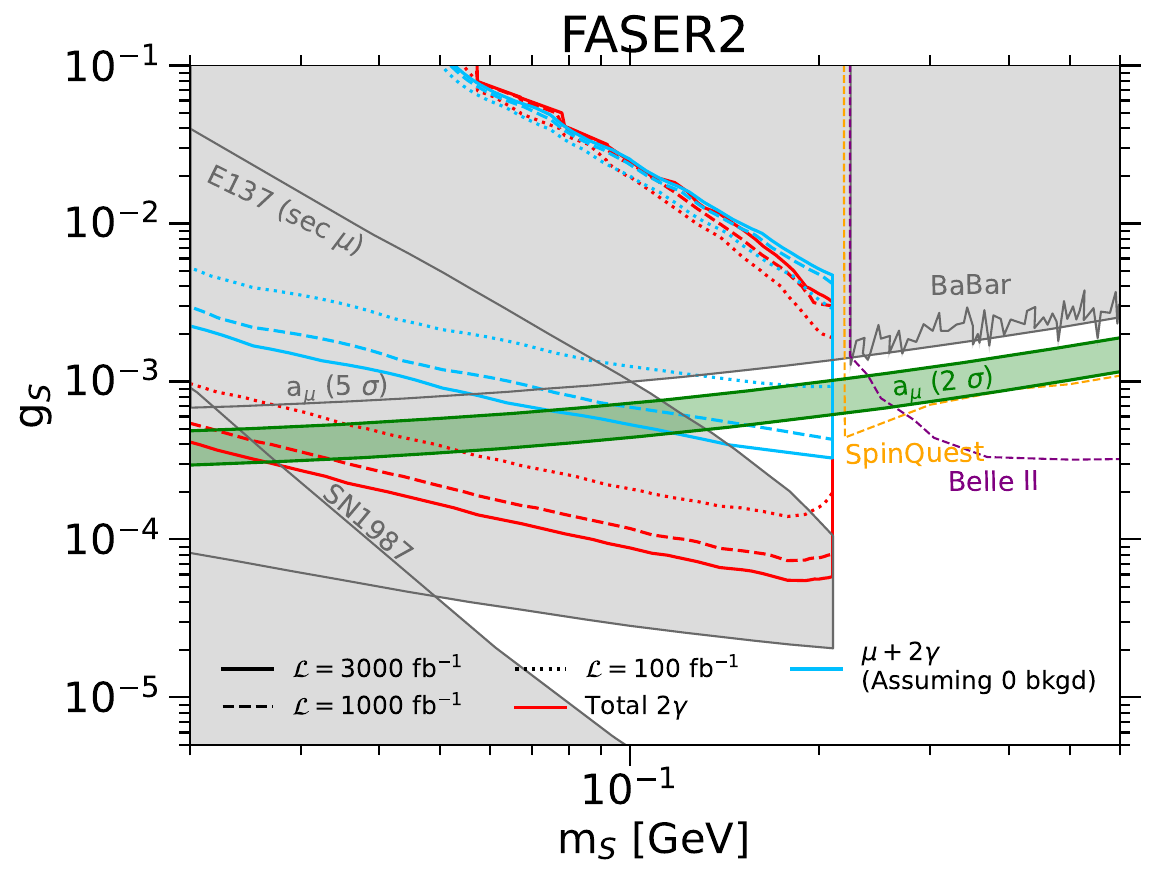}  
\includegraphics[width=0.48\textwidth]{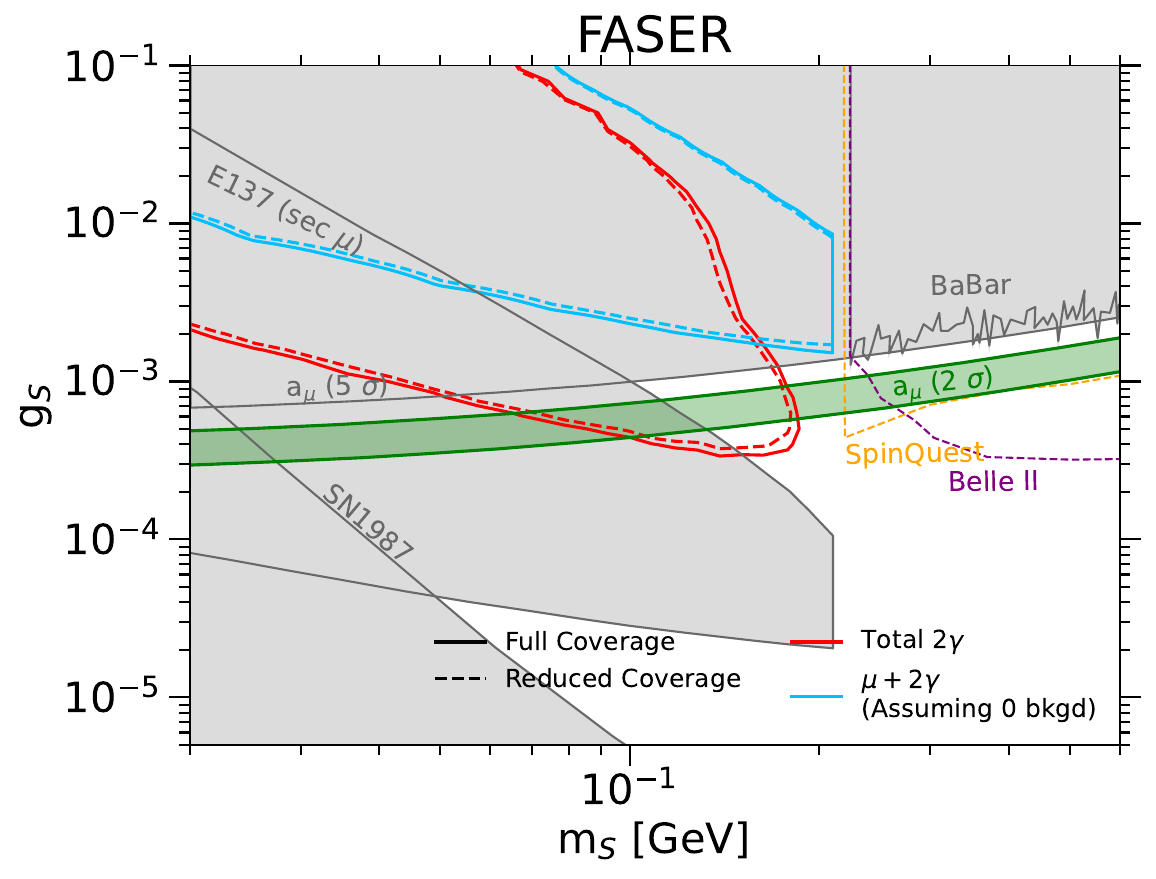} 
\includegraphics[width=0.48\textwidth]{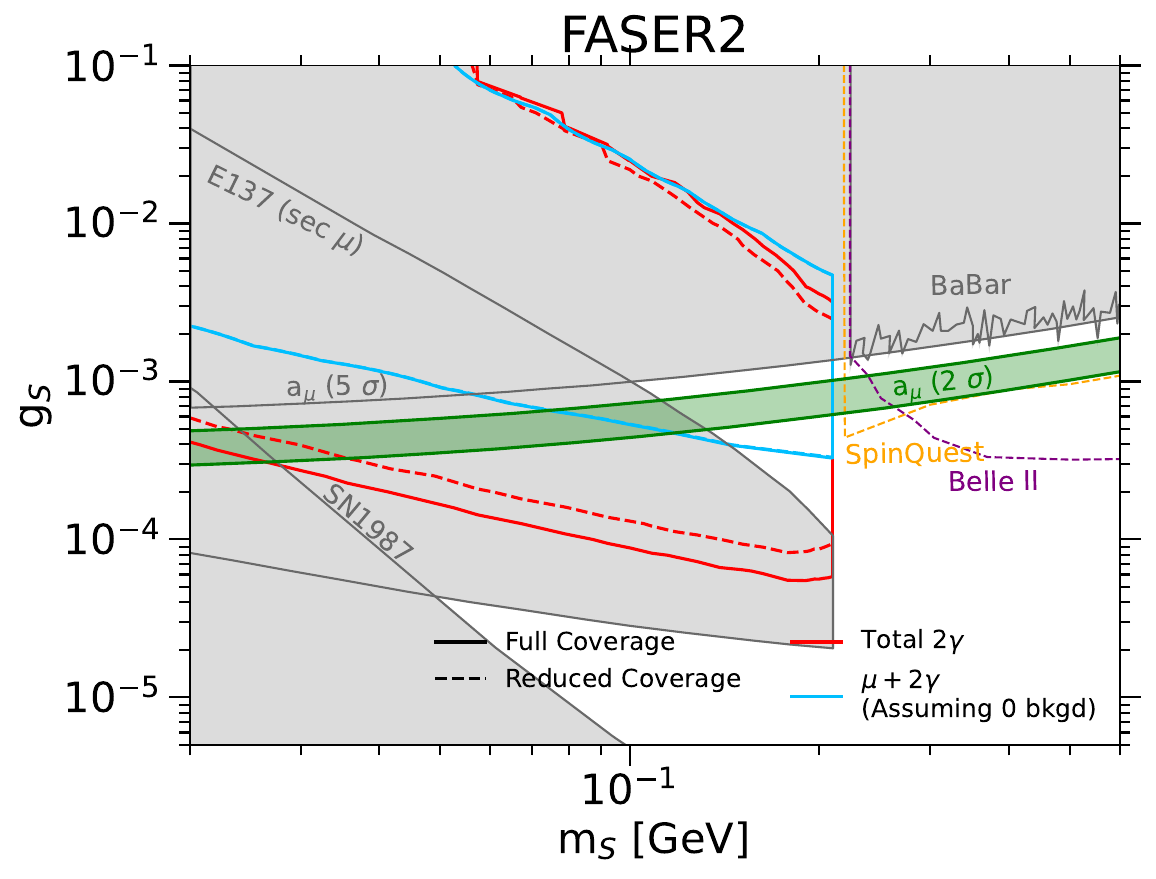}  
\caption{Projected exclusion at 90\% C.L. for FASER and FASER2.
\textbf{Top Row:}~The upper left panel shows projections 
for FASER (770 fb$^{-1}$), while the upper right panel shows it for FASER2 (3 ab$^{-1}$).
We display the summed contour lines for diphoton signal events from rare meson decays, Bremsstrahlung production in the TA(X)S, and rock.
Both panels require $\Delta_{\gamma\gamma}>200~\mu$m for background mitigation while varying $E_{\gamma\gamma}>$ \{0, 200, 500, 1000, 1500, 2000\} GeV. \textbf{Middle Row:}~The middle left and middle right panels show the exclusion contours for FASER and FASER2, respectively, for various luminosities, with cuts on the diphoton signal ($\Delta_{\gamma\gamma}>200~\mu$m and E${\gamma\gamma}>200$ GeV). Here we also show the reach for the muon + diphoton signal coming from Bremsstrahlung production in FASER$\nu$(2). The decreasing luminosities can be re-interpreted as contour lines for increasing number of events: \{3, 4.6, 11.5\} for FASER and \{3, 9, 90\} for FASER2.
\textbf{Bottom Row:}~Projected exclusion at FASER (770 fb$^{-1}$) and FASER2 (3 ab$^{-1}$) with full and reduced transverse area coverage by the high-granularity preshower. Similar to the middle panel, we show the summed contour lines for diphoton signal and the muon + diphoton signal events, with the same cuts ($E_{\gamma\gamma}>200$ GeV, $\Delta_{\gamma\gamma}>200~\mu$m) placed on the diphoton. 
In both panels, the full coverage corresponds to the upgraded preshower covering the full radius of the spectrometer, corresponding to a radius of $10~{\rm cm}$ ($1~\text{m}$). The reduced area corresponds to a radius of 8.6~cm (28~cm) centered around the middle of the detector.
In all the panels, the green band shows the preferred parameter space to solve the $(g-2)_{\mu}$ anomaly. Existing constraints are shown in gray.
} \label{fig:merged_reach_appendix} \end{figure*}

To fully characterize the discovery potential for this model, we show the exclusion reach at FASER and FASER2, for various kinematic cuts, luminosities, and preshower coverage.
In the top row of \cref{fig:merged_reach_appendix} we present the reach for various E$_{\gamma\gamma}$ cuts, while retaining the requirement of $\Delta_{\gamma\gamma}>200~\mu$m. Here we only show it for the diphoton signal events from rare meson
decays, Bremsstrahlung production in the TA(X)S and rock.
We see that even with stringent energy requirements of E$_{\gamma\gamma}>$ 500 GeV (2000 GeV), FASER(2) can probe open regions of the parameter space that can explain the $(g-2)_{\mu}$ anomaly.

In the middle row of \cref{fig:merged_reach_appendix}, we show the reach for various luminosities requiring $\Delta_{\gamma\gamma}>200~\mu$m and E$_{\gamma\gamma}>200$ GeV, for the diphoton signal events in red.
For FASER2, even with just 100 fb$^{-1}$ of data the reach can be saturated at the $m_S<2m_{\mu}$ limit. Note that the different luminosity curves can be reinterpreted as different number of signal event contours, in the case that unexpected backgrounds arise which require more than 3 events to qualify as an exclusion. To this effect, we also show the exclusion reach for the muon + diphoton signal in light-blue, which may encounter important backgrounds.

The installation of the high-granularity preshower is resource intensive, and thus the final implementation may not cover the entire FASER(2) 10cm (1m) aperture; so we also consider the sensitivity to our signal if only the inner 8.6cm (28cm) is covered, corresponding to an area of about 232cm$^2$ (2460 cm$^2$). We require that both signal photons be within the reduced area of the preshower. In the bottom row of \cref{fig:merged_reach_appendix}, we compare the projected exclusion reach for the reduced transverse coverage, compared to the full coverage where the latter is defined as the the preshower covering the full radius of the spectrometer. We find that with the reduced preshower coverage, FASER is still sensitive to unprobed regions of parameter space, in particular the $(g-2)_{\mu}$ region.


\bibliography{references}

\end{document}